\documentclass[prl, nofootinbib, aps, twocolumn, superscriptaddress,
showpacs, preprintnumbers, amsmath, amssymb, floatfix]{revtex4}

\usepackage{graphicx}

\newcommand{\lsim}{\buildrel<\over{_\sim}}
\newcommand{\gsim}{\buildrel>\over{_\sim}}
\newcommand{\gtsim}{\gtrsim}

\newcommand{\Order}{{\cal O}}   
\newcommand{\eV}{\mathrm{eV}}
\newcommand{\keV}{\mathrm{keV}}

\newcommand{\GeV}{\mathrm{GeV}}
\newcommand{\TeV}{\mathrm{TeV}}
\newcommand{\Mpc}{\mathrm{Mpc}}
\newcommand{\km}{\mathrm{km}}
\newcommand{\seconds}{\mathrm{s}}

\newcommand{\MPl}{\mathrm{M}_{\mathrm{P}}}
\newcommand{\gravitino}{{\widetilde{G}}}

\newcommand{\stau}{{\widetilde{\tau}_1}}

\newcommand{\gluino}{{\widetilde g}}
\newcommand{\slepton}{{\widetilde{l}_1}}
\newcommand{\mslepton}{m_{\slepton}}

\newcommand{\mgr}{m_{\widetilde{G}}}

\newcommand{\CDM}{\mathrm{DM}}

\newcommand{\MGUT}{M_{\mathrm{GUT}}}
\newcommand{\Reheating}{\mathrm{R}}
\newcommand{\TR}{T_{\Reheating}}

\newcommand{\Lisix}{{}^6 \mathrm{Li}}
\newcommand{\Hefour}{{}^4 \mathrm{He}}

\newcommand{\champ}{X^{\! -}}
\newcommand{\deuterium}{\mathrm{D}}

\newcommand{\Omegatp}{\Omega_{\widetilde{G}}^{\mathrm{TP}}}
\newcommand{\Omegantp}{\Omega_{\widetilde{G}}^{\mathrm{NTP}}}

\newcommand{\CBBN}{\mathrm{CBBN}}
\newcommand{\taumax}{\tau^{\max}}
\newcommand{\cmax}{c^{\max}}

\newcommand{\mgrmax}{m_{\gravitino}^{\max}}
\newcommand{\TRmax}{T_{\Reheating}^{\max}}
\newcommand{\OmegaGr}{\Omega_{\widetilde{G}}}
\newcommand{\OmegaDM}{\Omega_{\mathrm{DM}}}
\newcommand{\LiHprim}{^6\mathrm{Li/H}|_{\mathrm{p}}}

\begin{document}
%
%
\preprint{arXiv:0806.3266}
\preprint{MPP--2008--58}
%
%
%
\title{Probing the Reheating Temperature at Colliders and with Primordial Nucleosynthesis}
\author{Frank Daniel Steffen}
\email{steffen@mppmu.mpg.de}
\affiliation{Max-Planck-Institut f\"ur Physik, 
F\"ohringer Ring 6,
D--80805 Munich, Germany}
%
%
%
\begin{abstract}
  Considering gravitino dark matter scenarios with a long-lived
  charged slepton, we show that collider measurements of the slepton
  mass and its lifetime can probe not only the gravitino mass but also
  the post-inflationary reheating temperature $\TR$.
  In a model independent way, we derive upper limits on $\TR$ and
  discuss them in light of the constraints from the primordial
  catalysis of $^6$Li through bound-state effects.
  In the collider-friendly region of slepton masses below 1~TeV, the
  obtained conservative estimate of the maximum reheating temperature
  is about $\TR=3\times 10^9\,\GeV$
  for the limiting case of a small gluino--slepton mass splitting
  and about $\TR=10^8\,\GeV$
  for the case that is typical for universal soft supersymmetry
  breaking parameters at the scale of grand unification.
  We find that a determination of the gluino--slepton mass ratio at
  the Large Hadron Collider will test the possibility of $\TR>
  10^9\,\GeV$ and thereby the viability of thermal leptogenesis with
  hierarchical heavy right--handed Majorana neutrinos.
\end{abstract}
%
%
\maketitle
%
\section{Introduction}

Big bang nucleosynthesis (BBN) is a powerful tool to test physics
beyond the Standard Model;
cf.~\cite{Dimopoulos:1988ue,Jedamzik:2004er,Kawasaki:2004yh,Kawasaki:2004qu,Kohri:2005wn,Jedamzik:2006xz,Kawasaki:2008qe}
and references therein. Indeed, in supersymmetric (SUSY) theories,
severe constraints appear owing to the existence of the gravitino
$\gravitino$ which is the gauge field of local SUSY transformations
and whose mass is governed by the SUSY breaking scale.
As the spin-3/2 superpartner of the graviton, the gravitino is an
extremely weakly interacting particle with couplings suppressed by
inverse powers of the (reduced) Planck scale $\MPl=2.4\times
10^{18}\,\GeV$~\cite{Wess:1992cp}.  Accordingly, once produced in
thermal scattering of particles in the hot primordial
plasma~\cite{Ellis:1984eq,Moroi:1993mb,Ellis:1995mr,Bolz:1998ek,Bolz:2000fu,Pradler:2006qh,Pradler:2006hh,Rychkov:2007uq},
unstable gravitinos with a mass $\mgr \lsim 5~\TeV$ have long
lifetimes, $\tau_{\gravitino} \gsim 100~\seconds$, and decay during or
after BBN. Since the decay products affect the abundances of the
primordial light elements, successful BBN predictions imply a bound on
the reheating temperature after inflation $T_{\Reheating}$ which
governs the abundance of gravitinos before their
decay~\cite{Ellis:1984eq,Moroi:1993mb,Ellis:1995mr,Bolz:1998ek,Bolz:2000fu,Pradler:2006qh,Pradler:2006hh,Pradler:2007ne,Rychkov:2007uq}:
$T_{\Reheating}\lsim 10^8\,\GeV$ for $\mgr \lsim
5~\TeV$~\cite{Kohri:2005wn,Kawasaki:2008qe}.

We consider SUSY extensions of the Standard Model in which the
gravitino $\gravitino$ is the lightest supersymmetric particle (LSP)
and a charged slepton $\slepton$---such as the lighter stau
$\stau$---the next-to-lightest supersymmetric particle (NLSP).
Assuming R-parity conservation, the gravitino LSP is stable and a
promising candidate for dark matter;
cf.~\cite{Moroi:1993mb,Bolz:1998ek,Asaka:2000zh,Bolz:2000fu,Ellis:2003dn,Feng:2004mt,Cerdeno:2005eu,Steffen:2006hw,Pradler:2006qh,Pradler:2006hh,Pradler:2007ne,Rychkov:2007uq,Steffen:2007sp}
and references therein.  Because of the extremely weak interactions of
the gravitino, the NLSP typically has a long lifetime before it decays
into the gravitino. If these decays occur during or after BBN, the
Standard Model particles emitted in addition to the gravitino can
affect the abundances of the primordial light elements.  For the
charged slepton NLSP case, the BBN constraints associated with
hadronic/electromagnetic energy injection have been
estimated~\cite{Ellis:2003dn,Feng:2004mt,Cerdeno:2005eu,Steffen:2006hw}.
Taking into account additional constraints from large-scale-structure
formation, apparently viable gravitino dark matter scenarios had been
identified (see, e.g., benchmark scenarios $A_{1,2}$--$C_{1,2}$ in
Ref.~\cite{Steffen:2006hw}) which are attractive for two reasons:
(i)~$T_{\Reheating} > 10^9\,\GeV$ is possible so that thermal
leptogenesis with hierarchical right-handed
neutrinos~\cite{Fukugita:1986hr,Davidson:2002qv,Buchmuller:2004nz,Blanchet:2006be,Antusch:2006gy}%
\footnote{Note that flavor
  effects~\cite{Nardi:2006fx,Abada:2006fw,Blanchet:2006be,Antusch:2006gy}
  do not change the lower bound $T_{\Reheating} > 10^9\,\GeV$ required
  by successful thermal leptogenesis with hierarchical right-handed
  neutrinos~\cite{Blanchet:2006be,Antusch:2006gy}. However, in the
  case of (nearly) mass-degenerate heavy right-handed Majorana
  neutrinos, resonant leptogenesis can explain the baryon asymmetry at
  smaller values of
  $T_{\Reheating}$~\cite{Flanz:1994yx,Covi:1996fm,Pilaftsis:1997jf,Anisimov:2005hr}.
  Another example for a framework in which the limit $T_{\Reheating} >
  10^9\,\GeV$ is relaxed is non-thermal leptogenesis; see, e.g.,
  \cite{HahnWoernle:2008pq} and references therein.}
remains a viable explanation of the baryon
asymmetry~\cite{Bolz:1998ek,Fujii:2003nr,Cerdeno:2005eu,Steffen:2006hw,Pradler:2006qh}
and (ii)~the gravitino mass can be close to the NLSP mass $\mslepton$,
i.e., $0.1\,\mslepton\lesssim\mgr<\mslepton$, so that a kinematical
$\mgr$ determination appears
viable~\cite{Buchmuller:2004rq,Martyn:2006as,Hamaguchi:2006vu}. With a
kinematically determined $\mgr$, one would be able to measure the
Planck scale $\MPl$ at
colliders~\cite{Buchmuller:2004rq,Martyn:2006as,Hamaguchi:2006vu} and
to test the viability of thermal leptogenesis in the
laboratory~\cite{Pradler:2006qh}. Indeed, an agreement of the $\MPl$
value determined in collider experiments with the one inferred from
Newton's constant~\cite{Yao:2006px}
$G_{\rm N} = 6.709\times 10^{-39}\,\GeV^{-2}$
would provide evidence for the existence of supergravity (SUGRA) in
nature~\cite{Buchmuller:2004rq}.

Already in the early paper~\cite{Dimopoulos:1989hk} it had been
realized that long-lived negatively charged massive particles $\champ$
(such as long-lived $\slepton^-$'s) can form primordial bound states,
which can affect the abundances of the primordial light elements. It
was however only recently~\cite{Pospelov:2006sc} when it was realized
that bound-state formation of $\champ$ with $\Hefour$ can lead to a
substantial overproduction of primordial $\Lisix$ via the catalyzed
BBN (CBBN) reaction
\begin{equation}
(\Hefour\champ)+\deuterium \rightarrow \Lisix + \champ
\, .
\label{Eq:CBBNof6Li}
\end{equation}
With an $\champ$ abundance that is typical for an electrically charged
massive thermal relic~\cite{Asaka:2000zh}, this reaction becomes so
efficient that an $\champ$ lifetime of $\tau_{\champ} \gtrsim 5\times
10^3~\seconds$ is excluded by observationally inferred values of the
primordial $\Lisix$ abundance~\cite{Pospelov:2006sc}.  In the
considered gravitino LSP scenarios, this bound applies directly to the
lifetime of the $\slepton$
NLSP~\cite{Pospelov:2006sc,Cyburt:2006uv,Steffen:2006wx,Pradler:2006hh,Hamaguchi:2007mp,Kawasaki:2007xb,Pradler:2007is,Pradler:2007ar,Steffen:2007sp,Kawasaki:2008qe}:
$\tau_{\slepton}\lesssim 5 \times 10^3~\seconds$.%
\footnote{While numerous other bound-state effects can affect the
  abundances of $^6$Li and the other primordial light
  elements~\cite{Kohri:2006cn,Kaplinghat:2006qr,Cyburt:2006uv,Bird:2007ge,Jedamzik:2007cp,Jedamzik:2007qk},
  the approximate $\tau_{\slepton}$ bound is found to be quite robust;
  cf.~\cite{Pospelov:MPP-2008-63}.}
This implies $\mgr\lesssim 0.1\,\mslepton$ in the collider-friendly
mass range of $\mslepton\lesssim 1~\TeV$. Accordingly, the region
$0.1\,\mslepton\lesssim\mgr<\mslepton$, in which the kinematical
$\mgr$ determination appears feasible, seems to be excluded by BBN
constraints~\cite{Steffen:2006wx}.  Moreover, within the framework of
the constrained minimal supersymmetric Standard Model (CMSSM), we have
found that the $\Lisix$ constraint implies the upper limit $\TR
\lesssim 10^7~\GeV$ in gravitino dark matter scenarios with unbroken
R-parity and typical thermal $\stau$ NLSP relic
abundances~\cite{Pradler:2006hh,Pradler:2007is,Pradler:2007ar}. For a
standard cosmological history, this finding clearly disfavors
successful thermal leptogenesis within the CMSSM in the case of
hierarchical heavy right-handed Majorana neutrinos.  For a gravitino
LSP mass range that is natural for gravity-mediated SUSY breaking, the
$\Lisix$ constraint can even point to a CMSSM mass spectrum which will
be difficult to probe at the Large Hadron Collider
(LHC)~\cite{Cyburt:2006uv,Pradler:2006hh,Pradler:2007is,Pradler:2007ar}.

This letter provides a model independent study of gravitino LSP
scenarios with a charged slepton NLSP that has a $^6$Li-friendly
lifetime of $\tau_{\slepton}\lesssim 10^4~\seconds$ and a
collider-friendly mass of $\mslepton\lesssim 1~\TeV$. In contrast
to~\cite{Pradler:2006hh,Pradler:2007is,Pradler:2007ar,Kersten:2007ab},
the investigation presented in this work is not restricted to a
constrained framework such as the CMSSM or to the gravitino mass range
suggested by gravity-mediated SUSY breaking. Thereby, generic results
are obtained with a range of validity that includes models with
gauge-mediated SUSY breaking and/or non-standard mass spectra.

The remainder of this letter is organized as follows. 
In the next section we review that $\mgr$ could be determined by
measuring $\mslepton$ and $\tau_{\slepton}$ at colliders.
We give also an upper limit on $\mgr$ that depends on $\mslepton$ and
$\tau_{\slepton}$.
This limit allows us to derive a lower limit for the gravitino density
$\OmegaGr$. Since $\OmegaGr$ cannot exceed the dark matter density
$\OmegaDM$, this leads to con\-servative $\TR$ limits which can be
probed in measurements of $\mslepton$ and $\tau_{\slepton}$ (and the
gluino mass $m_{\gluino}$).
We use the derived expressions to translate the $\tau_{\slepton}$
constraint from CBBN of $^6$Li into robust upper limits on $\TR$ that
depend on the mass ratio $m_{\gluino}/\mslepton$.
Finally, we show that the requirement $\TR> 10^9\,\GeV$ needed for
successful standard thermal leptogenesis provides an upper limit on
this ratio and thereby a testable prediction for LHC phenomenology.

\section{Probing \boldmath$\mgr$ at Colliders}
\label{Sec:mGravitinoCollider}

In the considered SUSY scenarios with the gravitino LSP being stable
due to R-parity conservation,%
\footnote{For the case of broken $\mathrm{R}$-parity, see, e.g.,
  \cite{Buchmuller:2007ui}.}
the charged slepton NLSP has a lifetime $\tau_{\slepton}$ that is
governed by the decay $\slepton\to\gravitino\tau$ and thus given by
the SUGRA prediction for the associated partial width
\begin{equation}
        \tau_{\slepton} 
        \simeq \frac{48 \pi \mgr^2 \MPl^2}{\mslepton^5} 
        \left(1-\frac{\mgr^2}{\mslepton^2}\right)^{-4}
        \gtrsim \frac{48 \pi \mgr^2 \MPl^2}{\mslepton^5} 
        \, ,
\label{Eq:SleptonLifetime}
\end{equation}
where the rightmost term underestimates $\tau_{\slepton}$ by at most
5\% (30\%) for $\mgr \lesssim 0.1\,\mslepton$ ($\mgr \lesssim
0.25\,\mslepton$). 
Accordingly, using $\MPl=2.4\times 10^{18}\,\GeV$ as inferred from
Newton's constant, the gravitino mass $\mgr$ can be determined by
measuring both $\mslepton$ and $\tau_{\slepton}$ at future
colliders~\cite{Ambrosanio:2000ik}. Moreover, from the rightmost
expression in~(\ref{Eq:SleptonLifetime}), one can extract an upper
limit for the gravitino mass
%
%
\begin{equation}
        \mgr 
        \lesssim 
     0.41~\GeV
     \left(\frac{\tau_{\slepton}}{10^4~\seconds}\right)^{\!\!\frac{1}{2}}\!\!
     \left(\frac{\mslepton}{100~\GeV}\right)^{\!\!\frac{5}{2}}\!\!
        \equiv
        \mgrmax
        \, ,
\label{Eq:mGmax}
\end{equation}
which turns into an equality for $\mgr\ll\mslepton$. 

\begin{figure}
\includegraphics[width=3.25in]{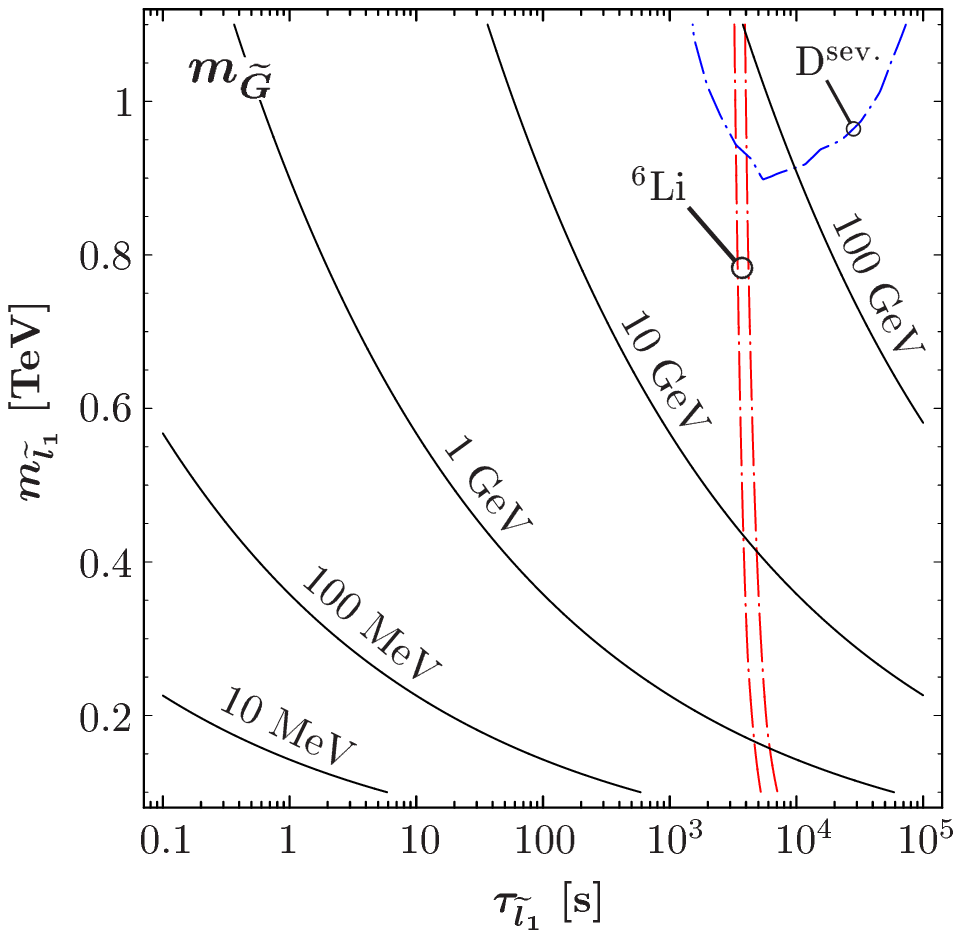}
\caption{Contours of $\mgr^{(\max)}$
  (solid lines) in the plane spanned by $\tau_{\slepton}$ and
  $\mslepton$. BBN constraints from D and $^6$Li are indicated by the
  short-dash-dotted (blue in the web version) and by the
  long-dash-dotted (red in the web version) lines, respectively.  The
  $^6$Li constraints are obtained from the CBBN treatment
  of~\cite{Pradler:2007is} for upper limits on the primordial
  $^6\mathrm{Li/H}$ abundance of $2\times 10^{-11}$ (left line) and
  $6\times 10^{-11}$ (right line).}
\label{Fig:ProbingMgravitino}
\end{figure}

In Fig.~\ref{Fig:ProbingMgravitino}
%
%
contours of $\mgr^{(\max)}$ (solid lines) are shown in the plane
spanned by $\tau_{\slepton}$ and $\mslepton$.
Only $m_{\slepton} \gtsim 100\,\GeV$ is considered since long-lived
charged sleptons should have otherwise been observed already at the
Large Electron--Positron Collider (LEP)~\cite{Yao:2006px}.

The potential for collider measurements of $\mslepton$ is promising
since each heavier superpartner produced will cascade down to the
$\slepton$ NLSP which will appear as a (quasi-) stable muon-like
particle in the
detector~\cite{Drees:1990yw,Nisati:1997gb,Feng:1997zr}. For slow
$\slepton$'s, the associated highly ionizing tracks and
time--of--flight measurements can allow for a distinction from
muons~\cite{Drees:1990yw,Nisati:1997gb,Feng:1997zr,Ambrosanio:2000ik}.
With measurements of the $\slepton$ velocity $\beta_{\slepton} \equiv
v_{\slepton}/c$ and its momentum $p_{\slepton}\equiv
|\vec{p}_{\slepton}|$, $\mslepton$ can be determined:
$\mslepton=p_{\slepton}(1-\beta_{\slepton}^2)^{1/2}/\beta_{\slepton}$~\cite{Ambrosanio:2000ik}.
For the upcoming LHC experiments, studies of hypothetical scenarios
with long-lived charged particles are actively
pursued~\cite{Ellis:2006vu,Bressler:2007gk,Zalewski:2007up}.  In
Ref.~\cite{Ellis:2006vu}, for example, it is shown that one should be
able to measure the mass $\mslepton$ of a (quasi-) stable slepton
quite accurately at the LHC.

The experimental determination of $\tau_{\slepton}$ will be
substantially more difficult than the $\mslepton$ measurement. If some
of the sleptons decay already in the collider detectors, the
statistical method proposed in~\cite{Ambrosanio:2000ik} could allow
one to measure $\tau_{\slepton}$. Moreover, ways to stop and collect
charged long-lived particles for an analysis of their decays have been
proposed for the LHC and for the International Linear Collider
(ILC)~\cite{Goity:1993ih,Buchmuller:2004rq,Hamaguchi:2004df,Feng:2004yi,DeRoeck:2005bw,Martyn:2006as,Hamaguchi:2006vu,Cakir:2007xa,Martyn:2007mj}.
These challenging proposals could lead to a precise measurement of
$\tau_{\slepton}$.

In addition, these proposals could help to distinguish the case of the
gravitino LSP from the one of the axino
LSP~\cite{Brandenburg:2005he+X,Hamaguchi:2006vu}, which
is---as the fermionic superpartner of the axion---another
well-motivated dark matter
candidate~\cite{Covi:1999ty,Steffen:2007sp}.
Indeed, also the axino LSP can be produced thermally in the early
Universe~\cite{Covi:2001nw,Brandenburg:2004du} and can be associated
with a $\slepton$ NLSP with
$\Order(\mathrm{ms})\lesssim\tau_{\slepton}\lesssim\Order(\mathrm{days})$
\cite{Covi:2004rb,Brandenburg:2005he+X}.
In the axino LSP case, however, measurements of $\tau_{\slepton}$ and
$\mslepton$ will probe the Peccei--Quinn scale
$f_a$~\cite{Brandenburg:2005he+X} instead of $\mgr$.
Keeping in mind that the axino LSP could mimic the gravitino LSP at
colliders (at first sight), we assume in the remainder of this work
that it is the gravitino LSP scenario that is realized in nature.

\section{Probing \boldmath$\TR$ at Colliders}

Within a standard cosmological history, the gravitino LSP can be
produced in decays of scalar fields such as the
inflaton~\cite{Asaka:2006bv,Takahashi:2007tz,Endo:2007sz}, in thermal
scattering of particles in the primordial
plasma~\cite{Moroi:1993mb,Bolz:1998ek,Bolz:2000fu,Pradler:2006qh,Pradler:2006hh,Rychkov:2007uq},
and in NLSP
decays~\cite{Borgani:1996ag,Asaka:2000zh,Ellis:2003dn,Feng:2004mt}.
Focussing on the case $\mgr\gg 100~\eV$ in which gravitinos are never
in thermal equilibrium due to their extremely weak interactions,%
\footnote{In gauge-mediated SUSY breaking scenarios, light gravitinos
  can be viable thermal relics if their abundance is diluted by
  entropy production, which can result, for example, from decays of
  messenger
  fields~\cite{Baltz:2001rq,Fujii:2002fv,Fujii:2003iw,Lemoine:2005hu,Jedamzik:2005ir}.}
the thermally produced gravitino density $\Omegatp$ depends basically
linearly on $\TR$---cf.~(\ref{Eq:OmegaGrTPSU3}) below---and thus can
serve as a thermometer of the earliest moments of the
radiation-dominated
epoch~\cite{Moroi:1993mb,Bolz:1998ek,Bolz:2000fu,Pradler:2006qh,Pradler:2006hh,Rychkov:2007uq}.
In particular, this allows for the derivation of upper limits on $\TR$
since the relic gravitino density is bounded from above by the
observed dark matter density~\cite{Yao:2006px}:
\begin{equation}
\Omegatp h^2 \lesssim
\OmegaGr h^2\leq
\OmegaDM h^2\simeq 0.1
\, ,
\label{Eq:OmegaDM}
\end{equation}
where $h\simeq 0.7$ denotes the Hubble constant in units of
$100~\km\,\Mpc^{-1}\seconds^{-1}$.  

Aiming at a lower limit of $\OmegaGr$ to arrive at a truly
conservative upper limit on $\TR$, we do not take into account the
model dependent contributions from decays of scalar fields such as the
inflaton~\cite{Asaka:2006bv,Takahashi:2007tz,Endo:2007sz} or the ones
from NLSP decays~\cite{Borgani:1996ag,Asaka:2000zh,Feng:2004mt},%
\footnote{For $\TR$ limits obtained by taking into account
  contributions to $\OmegaGr$ from NLSP decays, see
  e.g.~\cite{Asaka:2000zh,Fujii:2003nr,Cerdeno:2005eu,Steffen:2006hw,Pradler:2006qh,Pradler:2006hh,Choi:2007rh,Pradler:2007ar}.}
which can be negligible anyhow, in particular, for $\mgr \lesssim
1~\GeV$.  Indeed, we focus on the SUSY QCD contribution to the
thermally produced gravitino density, as derived in the
gauge-invariant calculation of~\cite{Bolz:2000fu,Pradler:2006qh},
\begin{eqnarray}
  \Omegatp h^2 |_{\mathrm{SU(3)}_c}
  &\!\!\!\!=\!\!& 
  0.117\,g_s^2(\TR)\left[1+\frac{m_{\tilde{g}}^2(\TR)}{3\mgr^2}\right]
  \ln\left[\frac{1.271}{g_s(\TR)}\right]
\nonumber\\
        &&
        \times
        \left(\frac{\mgr}{100~\GeV}\right)
        \left(\frac{T_{\Reheating}}{10^{10}\,\GeV}\right)
\label{Eq:OmegaGrTPSU3}
\end{eqnarray}
with the strong coupling $g_s$ and the gluino mass $m_{\gluino}$ to be
evaluated at the scale given by $\TR$, i.e.,
$g_s(T_R)=[g_s^{-2}(M_Z)\!+\!3\ln(T_R/M_Z)/(8\pi^2)]^{-1/2}$ and
$m_{\gluino}(T_R)=[g_s(T_R)/g_s(M_Z)]^2 m_{\gluino}(M_Z)$, where
$g_s^2(M_Z)/(4\pi)=0.1172$ at $M_Z=91.188~\GeV$.

By discarding the electroweak contributions, $\Omegatp$ is
underestimated typically by 20\%--50\% depending on the size of the
gaugino masses in the electroweak
sector~\cite{Pradler:2006qh,Pradler:2006hh,Pradler:2007ne}.  Thus,
these additional contributions would tighten the $\TR$ limits derived
below only by a factor that is not smaller than about 2/3.

For high-temperatures,
$10^6\,\GeV\lesssim\TR\lesssim 10^{10}\,\GeV$, 
where~(\ref{Eq:OmegaGrTPSU3}) derived in the weak coupling limit $g_s
\ll 1$ is most reliable~\cite{Bolz:2000fu,Pradler:2006qh},
we do now derive a lower limit
\begin{equation}
\OmegaGr^{\min} h^2 \lesssim
\Omegatp h^2 |_{\mathrm{SU(3)}_c}
\label{Eq:OmegaGconsVSOmegaGrTPSU3}
\end{equation}
by manipulating expression~(\ref{Eq:OmegaGrTPSU3}) as follows:
\begin{itemize}

\item[(i)] We use the replacement
\begin{equation}
  \ln\left[\frac{1.271}{g_s(\TR)}\right] 
  \to \ln\left[\frac{1.271}{g_s(10^6\,\GeV)}\right]
  = 0.256
  \, .
\end{equation}
Thereby, $\Omegatp h^2 |_{\mathrm{SU(3)}_c}$ is underestimated at most
at $\TR=10^{10}\,\GeV$ and there by a factor of about 5/8. 

\item [(ii)] We use $g_s(10^{10}\,\GeV)=0.85$ in the numerator
  of~(\ref{Eq:OmegaGrTPSU3}) and to evolve $m_{\gluino}(\TR)$ to the
  weak scale.
  Thereby, $\Omegatp h^2 |_{\mathrm{SU(3)}_c}$ is underestimated at
  most at $\TR=10^{6}\,\GeV$ and there by a factor of about 2/5. 

\item[(iii)] We use the constant $c>1$ to parametrize
  $m_{\gluino}(M_Z)$ in terms of the mass of the slepton NLSP:
\begin{equation}
      m_{\gluino}(M_Z) = c\, m_{\slepton}
      .
\label{Eq:mgluinomslepton}
\end{equation}
For concreteness, we use $m_{\gluino}(M_Z)$ to represent the gluino
mass at the weak scale. While the running mass $m_{\gluino}$ decreases
when evolved to higher energy scales, we assume implicitly
$m_{\gluino}>m_{\slepton}$ (even in the limiting case $c\simeq 1$) at
least up to energy scales accessible at the LHC and thereby up to
temperatures well above the freeze-out temperature of the $\slepton$
NLSP.  For example, since $m_{\gluino}$ decreases by about 20\% when
evolved from $M_Z$ to $10~\TeV$, $\Omegatp h^2 |_{\mathrm{SU(3)}_c}$
can thereby be underestimated for $c\simeq 1$ by a factor of about
$(4/5)^2$.

\item[(iv)] We drop the contributions from the spin 3/2 components of
  the gravitino which are given by the term in~(\ref{Eq:OmegaGrTPSU3})
  that is independent of $m_{\tilde{g}}$. Focussing on $\mgr\lesssim
  0.1\,\mslepton$, the relative importance of the spin 1/2 components
  is minimal for $\mgr\simeq 0.1\,\mslepton$, $c\simeq 1$, and
  $\TR=10^{10}\,\GeV$, where the associated second term in the first
  bracket of~(\ref{Eq:OmegaGrTPSU3}) becomes about 8 so that $\Omegatp
  h^2 |_{\mathrm{SU(3)}_c}$ is underestimated without the spin 3/2
  term by a factor that is not smaller than about 8/9.

\end{itemize}
Accordingly, we find a guaranteed gravitino density of
\begin{equation}
  \OmegaGr^{\min} h^2
  = 0.174 \left(\frac{1~\GeV}{\mgr}\right)\!\!
  \left(\frac{c\,\mslepton}{100~\GeV}\right)^{\!\!2}\!\!
  \left(\frac{\TR}{10^{10}\,\GeV}\right)
  .
\label{Eq:OmegaGmin}
\end{equation}
For $\mgr\lesssim 0.1\,\mslepton$,
$10^8\,\GeV\,(10^9\,\GeV)\,\leq\TR\leq 10^{10}\,\GeV$ and
$c\,\mslepton>200~\GeV~(100~\GeV)$, this expression is associated with
$1.57 \lesssim \Omegatp|_{\mathrm{SU(3)}_c}/\OmegaGr^{\min}\lesssim
2.1$ ($1.9$).
Accordingly, the limits derived below may be considered to be `too
relaxed' (or `too conservative') by at least a factor of about 1.5.
Indeed, $\Omegatp|_{\mathrm{SU(3)}_c}/\OmegaGr^{\min}$ increases by
decreasing $\TR$ and/or $c\,\mslepton$ or by increasing $\mgr$ so that
the corresponding limits presented below will become even more
conservative.
For example, for $\TR=10^6\,\GeV$, $c\,\mslepton\simeq 100\,\GeV$, and
$\mgr\lesssim 0.05\,\mslepton$,
$\Omegatp|_{\mathrm{SU(3)}_c}/\OmegaGr^{\min}\simeq 2.4$ is
encountered so that the associated limits derived below may be
considered as `too relaxed' by at least a factor of about 2.4.

With $\mgrmax$ from~(\ref{Eq:mGmax}), expression~(\ref{Eq:OmegaGmin})
leads to a lower limit on $\OmegaGr$ in terms of $\mslepton$ and
$\tau_{\slepton}$:
\begin{equation}
  \OmegaGr h^2 
  \gtrsim 0.422 \, c^2
  \left(\frac{10^4\,\seconds}{\tau_{\slepton}}\right)^{\!\!\frac{1}{2}}\!\!
  \left(\frac{100~\GeV}{\mslepton}\right)^{\!\!\frac{1}{2}}\!\!
  \left(\frac{\TR}{10^{10}\,\GeV}\right)
  .
\label{Eq:OmegaGlimit}
\end{equation}
Comparing this lower limit with the dark matter
constraint~(\ref{Eq:OmegaDM}), we arrive at a conservative $\TR$ limit
that can be determined at colliders by measuring the slepton mass
$\mslepton$, its lifetime $\tau_{\slepton}$, and $c$ (or
the gluino mass $m_{\gluino}$):
\begin{eqnarray}
  \TR 
  &\leq& 
  \frac{2.37\times 10^9~\GeV}{c^{2}} 
  \left(\frac{\Omega_{\CDM}h^2}{0.1}\right)
  \nonumber\\
  &&\times
  \left(\frac{\tau_{\slepton}}{10^4~\seconds}\right)^{\!\!\frac{1}{2}}\!\!
  \left(\frac{\mslepton}{100~\GeV}\right)^{\!\!\frac{1}{2}}\!\!
  \equiv
  \TR^{\max}
  \, .
\label{Eq:TR_max}
\end{eqnarray}
This is one of the main results of this letter. Note that this limit
can be refined easily. Once an additional contribution
$\Omega_{\mathrm{x}}$ to $\Omega_{\CDM}$---such as an axion density or
a non-thermally produced gravitino density $\Omegantp$---is taken for
granted, the resulting tighter $\TR$ limits can be obtained from
(\ref{Eq:TR_max}) after the replacement:
$\OmegaDM\to\OmegaDM-\Omega_{\mathrm{x}}$.
In fact, based on the interplay between $\Omegatp$ and the
contribution from NLSP decays, $\Omegantp$-dependent upper limits on
$\TR$ have been derived to be tested at colliders~\cite{Choi:2007rh}.

At this point, one has to clarify to which $\TR$ definition the
limit~(\ref{Eq:TR_max}) applies.  The analytic
expression~(\ref{Eq:OmegaGrTPSU3}) is derived by assuming a
radiation-dominated epoch with an initial temperature of
$\TR$~\cite{Bolz:2000fu,Pradler:2006qh}.
In a numerical treatment, the epoch in which the coherent oscillations
of the inflaton field dominate the energy density of the Universe can
also be taken into account, where one usually defines $\TR$ in terms
of the decay width $\Gamma_{\phi}$ of the inflaton
field~\cite{Kawasaki:2004qu,Pradler:2006hh}. In fact, the numerical
result for $\Omegatp$ has been found to agree with the corresponding
analytic expression for~\cite{Pradler:2006hh}
\begin{equation}
  \TR 
  \simeq
  \left[\frac{90}{g_*(\TR)\pi^2}\right]^{1/4}
  \sqrt{\frac{\Gamma_{\phi}\MPl}{1.8}}
\label{Eq:TR_definition}
\end{equation}
which satisfies $\Gamma_{\phi}\simeq 1.8 H_{\mathrm{rad}}(\TR)$ with
the Hubble parameter 
$H_{\mathrm{rad}}(T)=\sqrt{g_*(T)\pi^2/90}\, T^2/\MPl$ 
and an effective number of relativistic degrees of freedom of
$g_*(\TR)=228.75$.
Thus, (\ref{Eq:TR_definition}) provides the $\TR$ definition to which
the upper limit~(\ref{Eq:TR_max}) applies. For an alternative $\TR$
definition given by $\Gamma_{\phi}=\xi H_{\mathrm{rad}}(\TR^{[\xi]})$,
\begin{equation}
  \TR^{[\xi]} 
  \equiv
  \left[\frac{90}{g_*(\TR)\pi^2}\right]^{1/4}
  \sqrt{\frac{\Gamma_{\phi}\MPl}{\xi}}
  \, ,
\label{Eq:TR_xi_definition}
\end{equation}
the upper limit~(\ref{Eq:TR_max}) can be translated accordingly
\begin{equation}
     \TR^{\max\,[\xi]}=\sqrt{\frac{1.8}{\xi}}\,\TR^{\max}
     \, .      
\label{Eq:TR_max_xi}
\end{equation}
In particular, the associated numerically obtained $\Omegatp$ can be
reproduced with the analytical expression after substituting $\TR$
with $\sqrt{\xi/1.8}\,\TR^{[\xi]}$~\cite{Pradler:2006hh}.

\begin{figure}
\includegraphics[width=3.25in]{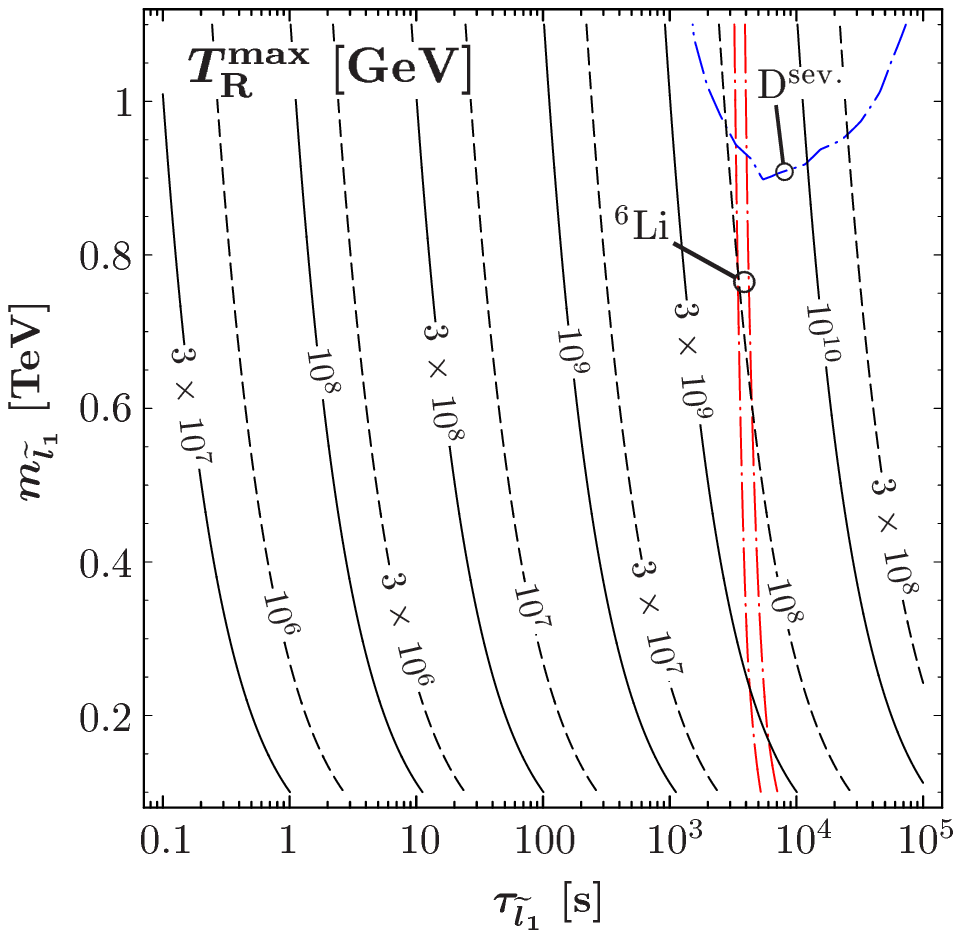}
\caption{Contours of $\TRmax$ imposed by $\OmegaGr h^2\leq\OmegaDM
  h^2\leq 0.126$ for $c=1$ (solid lines) and $c=7$ (dashed lines) in
  the plane spanned by $\tau_{\slepton}$ and $\mslepton$.  The BBN
  constraints are identical to the ones shown in
  Fig.~\ref{Fig:ProbingMgravitino}.}
\label{Fig:ProbingTRmax}
\end{figure}
%
In Fig.~\ref{Fig:ProbingTRmax} contours of the upper limit $\TRmax$
given in~(\ref{Eq:TR_max}) are shown for $c=1$ (solid lines) and $c=7$
(dashed lines) as obtained with
$\OmegaDM h^2\leq 0.126$.%
\footnote{This $\OmegaDM h^2$ value is used to allow for a convenient
  comparison with our previous
  works~\cite{Pradler:2006qh,Pradler:2006hh}. There $\OmegaDM h^2\leq
  0.126$ is understood as a nominal $3\sigma$ limit derived with a
  restrictive six-parameter ``vanilla'' model from the three year data
  set of the Wilkinson Microwave Anisotropy Probe (WMAP)
  satellite~\cite{Spergel:2006hy}.}
The most conservative upper limit is represented by the limiting case%
\footnote{This limiting case has already been considered in the
  benchmark scenarios $C_{1,2}$ in Ref.~\cite{Steffen:2006hw} which
  are now understood to be disfavored by the $^6$Li constraint from
  CBBN~\cite{Pospelov:2006sc}.}
$c=1$ which holds even without insights into $m_{\gluino}$.
Indeed, in this exceptional case, we assume implicitly a mass
difference, $\Delta m=m_{\gluino}-m_{\slepton}>0$, such that
$\slepton$--$\gluino$ coannihilation effects are negligible and such
that long-lived gluinos do not appear (avoiding thereby possibly
severe hadronic BBN constraints from late decaying
$\gluino$'s~\cite{Fujii:2003nr}).

With experimental insights into $m_{\slepton}$ and $m_{\gluino}$ (and
thereby into $c$), the limit~(\ref{Eq:TR_max}) can become considerably
more severe than in the $c=1$ case. In Fig.~\ref{Fig:ProbingTRmax}
this is illustrated for $c=7$ which is a value that appears typically
in constrained scenarios with universal soft SUSY breaking parameters
at the scale of grand unification $\MGUT\simeq 2\times 10^{16}\,\GeV$.
For example, in the CMSSM in which the gaugino masses, the scalar
masses, and the trilinear scalar couplings are assumed to take on the
respective universal values $m_{1/2}$, $m_0$, and $A_0$ at $\MGUT$,
the $\stau$ NLSP region has been found to be associated with
$m_{\stau}^2\leq 0.21\, m_{1/2}^2$ and thus with $c>6$ over the entire
natural parameter range~\cite{Pradler:2007is}.

The $\TR$ limits derived above are in spirit similar to the
$\mgr$-dependent $\TR$ limits given in
Refs.~\cite{Moroi:1993mb,Asaka:2000zh,Cerdeno:2005eu,Steffen:2006hw,Pradler:2006hh,Rychkov:2007uq,Choi:2007rh}.
Also the sensitivity of these $\TR$ limits on the gaugino masses has
already been discussed and used to provide $\mgr$-dependent upper
limits on the gaugino masses for given values of
$\TR$~\cite{Bolz:1998ek,Bolz:2000fu,Fujii:2003nr,Pradler:2006qh}.
In fact, once the gaugino masses and the gravitino mass $\mgr$
(inferred from $m_{\slepton}$ and $\tau_{\slepton}$) are known,
numerical results such as the ones shown in Fig.~2 of
Ref.~\cite{Pradler:2006hh} will provide a $\TR$ limit that is more
restrictive than our analytic estimate~(\ref{Eq:TR_max}). However, as
long as insights into a SUSY model possibly realized in nature are
missing, we find it important to provide a conservative and robust
estimate of $\TRmax$ that is insensitive to details in the SUSY
spectrum (other than the assumption of the $\gravitino$ LSP and a
$\slepton$ NLSP).  In particular, the comparison of~(\ref{Eq:TR_max})
with more model dependent limits---obtained, e.g., within the
CMSSM~\cite{Pradler:2006hh,Pradler:2007is,Pradler:2007ar}---demonstrates
very clearly the impact of restrictive assumptions on the soft SUSY
breaking sector.  Moreover, the derivation of $\TRmax$ contours as a
function of $\mslepton$ and $\tau_{\slepton}$ may turn out to become
very useful since $\mslepton$ and $\tau_{\slepton}$ are the quantities
that might be directly accessible in collider experiments. Finally, as
demonstrated below, the presentation of $\TRmax$ in the plane spanned
by $\tau_{\slepton}$ and $\mslepton$ allows also for a convenient
analysis of $\TRmax$ in light of recent BBN constraints.

Before proceeding, we would like to stress that the relation of
$\TRmax$ to quantities $\mslepton$, $\tau_{\slepton}$, and $c$ (or
$m_{\gluino}$), which could be accessible at future collider
experiments, relies crucially on assumptions on the cosmological
history and the evolution of physical parameters. 
For example, for a non-standard thermal history with late-time entropy
production, the thermally produced gravitino abundance can be diluted
$\Omegatp\to\Omegatp/\delta$ by a factor $\delta>1$ so that
$\TRmax\to\delta\,\TRmax$~\cite{Pradler:2006hh}.
Moreover, the limit~(\ref{Eq:TR_max}) can be evaded if the strong
coupling $g_s$ levels off in a non-standard way at high
temperatures~\cite{Buchmuller:2003is}.
This emphasizes that~(\ref{Eq:TR_max}) relies on the assumptions of a
standard cosmological history and a strong gauge coupling that behaves
at high temperatures as described by the renormalization group
equation in the minimal supersymmetric Standard Model (MSSM),
i.e., as described by $g_s(\TR)$ given below~(\ref{Eq:OmegaGrTPSU3}).
While tests of these assumptions seem inaccessible to terrestrial
accelerator experiments, futuristic space-based gravitational-wave
detectors such as the Big Bang Observer (BBO) or the Deci-hertz
Interferometer Gravitational Wave Observatory
(DECIGO)~\cite{Seto:2001qf} could allow for tests of the thermal
history after inflation and could even probe
$\TR$~\cite{Nakayama:2008ip,Nakayama:2008wy} in a way that is
complementary to the approach presented in this letter.

\section{Probing \boldmath$\TR$ with Primordial Nucleosynthesis}
\label{Sec:TRmaxBBN}

Gravitino LSP scenarios with a long-lived charged slepton NLSP can
affect BBN as described in the Introduction.  Particularly severe is
the catalytic effect of ($^4$He$\slepton^-$)-bound states on the
primordial abundance of $^6$Li~\cite{Pospelov:2006sc}. Indeed, the
CBBN reaction~(\ref{Eq:CBBNof6Li}) can become very efficient at
temperatures $T\simeq 10~\keV$ depending on the $\slepton$ abundance
at that time.  Observationally inferred upper limits on the primordial
$^6$Li/H abundance $\LiHprim$ (cf.~\cite{Asplund:2005yt}) can thus be
translated into $\tau_{\slepton}$-dependent upper limits on the
thermal relic abundance of the negatively charged $\slepton^-$
NLSP~\cite{Pospelov:2006sc,Hamaguchi:2007mp,Kawasaki:2007xb,Takayama:2007du,Pradler:2007is,Kawasaki:2008qe}.
For a given SUSY model, this abundance can be calculated, for example,
with the computer
program~\texttt{micrOMEGAs~2.1}~\cite{Belanger:2001fz+X}.  By
confronting the obtained abundance with the
$\tau_{\slepton}$-dependent upper limits 
(see, e.g., Fig.~1 in Ref.~\cite{Pradler:2007is}),
one can extract an upper limit on
$\tau_{\slepton}$~\cite{Pospelov:2006sc,Cyburt:2006uv,Steffen:2006wx,Pradler:2006hh,Hamaguchi:2007mp,Kawasaki:2007xb,Takayama:2007du,Pradler:2007is,Kawasaki:2008qe}.
This $\tau_{\slepton}$ limit can then be used to determine $\TRmax$
directly from~(\ref{Eq:TR_max}).

Let us perform this procedure explicitly for the following thermal
relic $\slepton$ NLSP abundance after decoupling and prior to
decay~\cite{Asaka:2000zh,Fujii:2003nr}:
\begin{equation}
  Y_{\slepton}
  \equiv \frac{n_{\slepton}}{s}
  = 2\,Y_{\slepton^-}
  = 0.7 \times 10^{-13}\left(\frac{\mslepton}{100~\GeV}\right)
  ,
\label{Eq:Yslepton}
\end{equation}
where $s$ denotes the entropy density and $n_{\slepton}$ the total
$\slepton$ number density assuming an equal number density of
positively and negatively charged $\slepton$'s.

We confront~(\ref{Eq:Yslepton}) with the $Y_{\slepton^-}$ limits that
emerge from a calculation of the $^6$Li abundance from CBBN,
$^6$Li/H$|_{\CBBN}$, which uses the state-of-the-art result of the
catalyzed $^6$Li production cross section~\cite{Hamaguchi:2007mp} and
the Boltzmann equation (instead of the Saha type approximation) to
describe the time evolution of the ($^4$He$\slepton^-$)-bound-state
abundance~\cite{Takayama:2007du,Pradler:2007is}.%
\footnote{We thank Josef Pradler for providing us with the
  $^6$Li/H$|_{\CBBN}$ data from the CBBN treatment of
  Ref.~\cite{Pradler:2007is}.}
Contour lines of $^6$Li/H$|_{\CBBN}$ obtained in this calculation are
shown, e.g., in Fig.~1 of Ref.~\cite{Pradler:2007is}.%
\footnote{Comparisons of~\cite{Pradler:2007is}
  with~\cite{Cyburt:2006uv,Kawasaki:2007xb,Jedamzik:2007qk,Kawasaki:2008qe},
  in which also the possible destruction of $^6$Li due to $\slepton$
  decays is considered, show that those effects affect the $^6$Li
  constraint only marginally.}
Working with an upper limit on the primordial $^6$Li/H abundance
of~\cite{Cyburt:2002uv}
\begin{equation}
        \LiHprim \leq 2\times 10^{-11}
        ,
\label{Eq:LiLimit1}
\end{equation}
the corresponding contour given in that figure is
the one that provides the upper limit for $Y_{\slepton^-}$.
As a second more conservative limit, we consider in this work also 
\begin{equation}
        \LiHprim \leq 6\times 10^{-11}
        ,
\label{Eq:LiLimit2}
\end{equation}
which provides a more relaxed $Y_{\slepton^-}$ limit.

In Figs.~\ref{Fig:ProbingMgravitino} and~\ref{Fig:ProbingTRmax}, the
long-dash-dotted (red in the web version) lines show the constraints
obtained by confronting~(\ref{Eq:Yslepton}) with the $Y_{\slepton^-}$
limits for $\LiHprim\leq 2\times 10^{-11}$ (left line) and $6\times
10^{-11}$ (right line).
It is the region to the right of the respective line that is
disfavored by $^6$Li/H$|_{\CBBN}$ exceeding the limit inferred from
observations.
Only the constraint from the primordial D abundance on hadronic energy
release~\cite{Kawasaki:2004qu,Jedamzik:2006xz} in $\slepton$
decays~\cite{Feng:2004mt,Cerdeno:2005eu,Steffen:2006hw} can be more
severe than the CBBN
constraints~\cite{Cyburt:2006uv,Steffen:2006wx,Pradler:2006hh,Kawasaki:2007xb,Steffen:2007sp,Kawasaki:2008qe}.%
\footnote{The additional primordial bound-state effects discussed
  in~\cite{Kohri:2006cn,Kaplinghat:2006qr,Cyburt:2006uv,Bird:2007ge,Kawasaki:2007xb,Jedamzik:2007cp,Jedamzik:2007qk,Pospelov:2007js}
  do not affect the conclusions of this letter;
  cf.~\cite{Pospelov:MPP-2008-63}.}
In Figs.~\ref{Fig:ProbingMgravitino} and~\ref{Fig:ProbingTRmax}, the
associated disfavored region is the one above the short-dash-dotted
(blue in the web version) line.  For details on this constraint,
see~\cite{Steffen:2006hw,Pradler:2006hh} and references therein.

In Fig.~\ref{Fig:ProbingMgravitino} one sees explicitly that the BBN
constraints disfavor the region
$0.1\,\mslepton\lesssim\mgr<\mslepton$ 
as already emphasized in the Introduction. Moreover, the $^6$Li
constraint imposes for $\mgr\gtrsim 10~\GeV$ the lower limit
$\mslepton>400~\GeV$ which implies $m_{\gluino}(M_Z)>2.4~\TeV$ for
$c>6$. This is in agreement with what had previously been realized in
the CMSSM (where $c>6$ in the $\stau$ NLSP
region~\cite{Pradler:2007is}):
For $\mgr$ in the range that is natural for gravity-mediated SUSY
breaking, the $^6$Li constraint can point to a SUSY mass spectrum that
will be difficult to probe at the
LHC~\cite{Cyburt:2006uv,Pradler:2006hh,Pradler:2007is,Pradler:2007ar}.
On the other hand, Fig.~\ref{Fig:ProbingMgravitino} shows very clearly
that the $^6$Li constraint becomes negligible for $\mgr < 1~\GeV$,
i.e., for the $\mgr$ range that is natural for gauge-mediated SUSY
breaking scenarios.

Figure~\ref{Fig:ProbingTRmax} allows us to read off the $\TRmax$ value
imposed by the $^6$Li constraint in the collider-friendly region of
$\mslepton\lesssim 1~\TeV$. Our most conservative and thereby most
robust limit is the one obtained in the limiting case $c=1$.
There one finds the $\TRmax=3\times 10^9\,\GeV$ contour in the region
that is allowed by the BBN constraints.
If the gluino turns out to be significantly heavier than the
$\slepton$ NLSP, the $\TRmax$ value will become considerably more
severe. For example, for the case of $c=7$, shown by the dashed lines
in Fig.~\ref{Fig:ProbingTRmax}, one finds that a reheating temperature
above $\TRmax=10^8\,\GeV$ is disfavored by BBN constraints.
Note that these limit are quite robust since they emerge from the
conservative limit~(\ref{Eq:TR_max}). In fact, one may consider these
limits as overly conservative by at least a factor of about 1.5 as
discussed below~(\ref{Eq:OmegaGmin}).
Indeed, a more restrictive limit of $\TR \lesssim 10^7\,\GeV$ is found
within the CMSSM by using the full expressions for $\Omegatp$ given
in~\cite{Pradler:2006qh} which include also the electroweak
contributions to thermal gravitino
production~\cite{Pradler:2006hh,Pradler:2007is,Pradler:2007ar}.

Let us comment on the model dependence of the BBN constraints. As
described above, the BBN constraints shown in
Figs.~\ref{Fig:ProbingMgravitino} and~\ref{Fig:ProbingTRmax} are
derived with the yield~(\ref{Eq:Yslepton}). This does introduce a
model dependence into the $\TRmax$ values discussed in the preceding
paragraph. However, one should stress two points: (i)~The
yield~(\ref{Eq:Yslepton}) is quite typical for an electrically charged
massive thermal relic~\cite{Asaka:2000zh}. (ii)~The
$\tau_{\slepton}$-dependent upper limits on $Y_{\slepton^-}$ are very
steep in the relevant region as can be seen, e.g., in Fig.~1 of
Ref.~\cite{Pradler:2007is}.
Indeed, this steepness is reflected by the relatively weak
$m_{\slepton}$ dependence of the $^6$Li constraints that can be seen in
Figs.~\ref{Fig:ProbingMgravitino} and~\ref{Fig:ProbingTRmax}.
These two points support that the shown BBN constraints and the
associated $\TRmax$ values are quite robust. Only for very generous
upper limits on the primordial $^6$Li abundance and/or exceptionally
small $Y_{\slepton^-}$ values can the CBBN bound be relaxed
substantially. The latter could be achieved, for example, with
non-standard entropy production after thermal freeze out of the
$\slepton$ NLSP and before
BBN~\cite{Buchmuller:2006tt,Pradler:2006hh,Hamaguchi:2007mp} which we
do not consider since we assume a standard cosmological history
throughout this letter.

\section{Probing \boldmath$\TR$ by Measuring the Gluino--Slepton Mass Ratio}

In this section we present a way that could allow us to probe with
collider experiments the $\TRmax$ value imposed by the $^6$Li
constraint---or any other constraint that can be translated into a
similar upper limit on $\tau_{\slepton}$---and the dark matter
constraint~(\ref{Eq:OmegaDM}).
As discussed above, the $^6$Li constraint implies basically an upper
limit on the $\slepton$ NLSP lifetime
\begin{equation}
        \tau_{\slepton} \leq \taumax,
\label{Eq:taumax}
\end{equation}
which is, e.g., $\taumax\simeq 5\times 10^3\,\seconds$ for the case of
the yield~(\ref{Eq:Yslepton}) as can be seen explicitly in
Fig.~\ref{Fig:ProbingTRmax}.
With a given $\taumax$ and~(\ref{Eq:TR_max}) derived from the dark
matter constraint~(\ref{Eq:OmegaDM}), one arrives immediately at the
following $\mgr$-independent upper limit on the gluino--slepton mass
ratio at the weak scale:
\begin{eqnarray}
  c 
  &\leq&
  \left(\frac{2.37\times 10^9~\GeV}{\TRmax}\right)^{\!\!\frac{1}{2}}\!\!
  \left(\frac{\Omega_{\CDM}h^2}{0.1}\right)^{\!\!\frac{1}{2}}\!\!
  \nonumber\\
  &&\times
  \left(\frac{\taumax}{10^4~\seconds}\right)^{\!\!\frac{1}{4}}\!\!
  \left(\frac{\mslepton}{100~\GeV}\right)^{\!\!\frac{1}{4}}\!\!
  \equiv \cmax,
\label{Eq:cmax}
\end{eqnarray}
i.e., a reheating temperature of at most $\TRmax$ can viable with a
given $\mslepton$ only for a mass ratio $c\leq\cmax$.
In other words, for a lifetime constraint $\taumax$ inferred from
cosmological considerations (such as CBBN) and/or collider
experiments, a simultaneous measurement of $\mslepton$ and the mass
ratio $c$ will provide a model-independent conservative limit $\TRmax$
for $\gravitino$ LSP scenarios with a $\slepton$ NLSP.

\begin{figure}
  \includegraphics[width=3.25in]{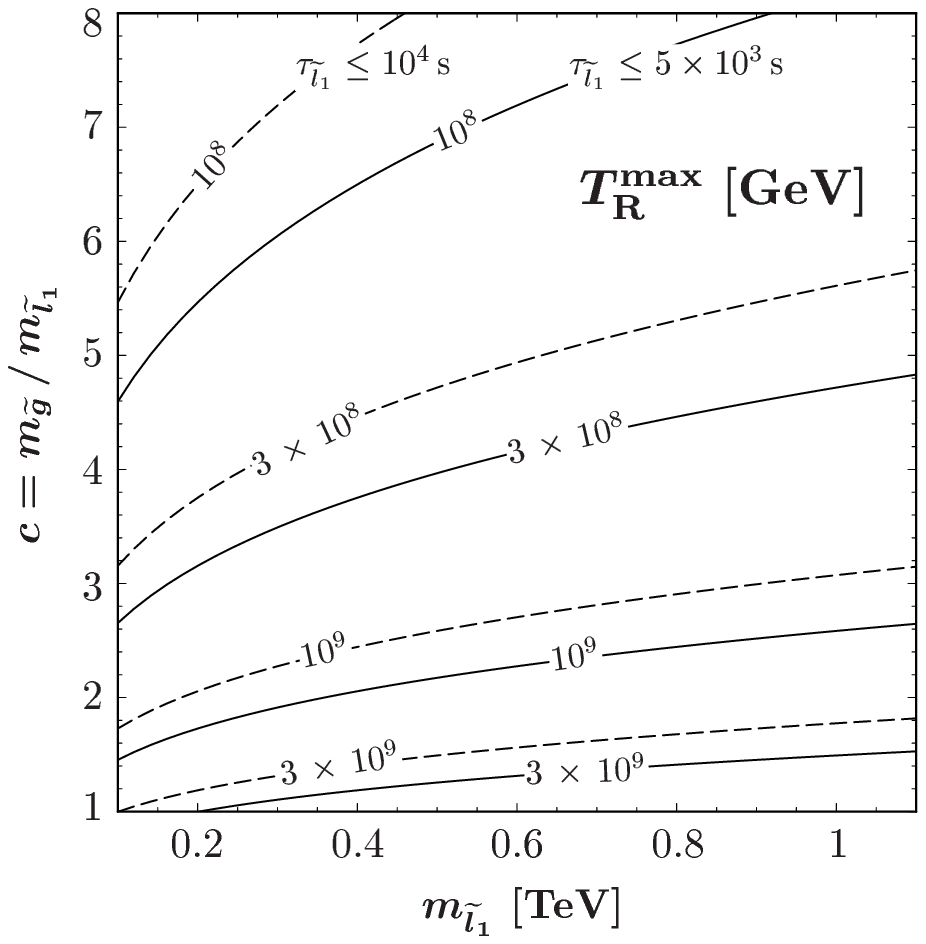}
  \caption{Upper limits on the mass ratio
    $c=m_{\gluino}(M_Z)/\mslepton$ imposed by $\OmegaGr
    h^2\leq\OmegaDM h^2\leq 0.126$ and $\tau_\slepton\leq3\times
    10^3\,\seconds$ ($10^4\,\seconds$) are shown as a function of
    $\mslepton$ by the solid lines (dashed lines) for values of
    $\TRmax$ ranging from $10^8\,\GeV$ up to $3\times 10^9\,\GeV$.}
\label{Fig:ProbingC}
\end{figure}

In Fig.~\ref{Fig:ProbingC}
%
%
the solid lines (dashed lines) show $\cmax$ imposed by the lifetime
constraint $\tau_{\slepton}\leq 3\times 10^3\,\seconds$
($10^4\,\seconds$) and and the dark matter constraint $\OmegaGr
h^2\leq\OmegaDM h^2\leq 0.126$ as a function of $\mslepton$ for values
of $\TRmax$ ranging from $10^8\,\GeV$ up to $3\times 10^9\,\GeV$.
The lifetime constraint $\tau_{\slepton}\leq 5\times 10^3\,\seconds$
is slightly weaker (and thereby more conservative) than the $^6$Li
CBBN constraint obtained with the yield~(\ref{Eq:Yslepton}) for
$\mslepton > 100~\GeV$ as can be seen in Fig.~\ref{Fig:ProbingTRmax}.
Accordingly, the $\TRmax$ values for $c=1$ and $c=7$ discussed in the
previous section are more restrictive than the corresponding $\TRmax$
values in Fig.~\ref{Fig:ProbingC}.
The even more conservative $\cmax$ curves obtained for
$\tau_{\slepton}\leq 10^4\,\seconds$ are presented to cover also the
case of a more relaxed CBBN constraint which could result, for
example, from $\LiHprim \leq 2.7\times
10^{-10}$~\cite{Jedamzik:2007qk}, i.e., an upper limit that is about
an order of magnitude more generous than~(\ref{Eq:LiLimit1}).
This emphasizes the conservative character of the corresponding
$\cmax$ curves shown in Fig.~\ref{Fig:ProbingC}.

\section{Testing the Viability of Thermal Leptogenesis at Colliders}

Thermal leptogenesis provides an attractive explanation of the baryon
asymmetry in the Universe~\cite{Fukugita:1986hr}. Since successful
thermal leptogenesis with hierarchical right-handed heavy Majorana
neutrinos requires a reheating temperature of
$\TR>10^9\,\GeV$~\cite{Davidson:2002qv,Buchmuller:2004nz,Blanchet:2006be,Antusch:2006gy},
one will be able to test its viability at the LHC with the method
described in the previous section.

From earlier studies of gravitino LSP
scenarios~\cite{Bolz:1998ek,Bolz:2000fu,Fujii:2003nr,Pradler:2006qh},
it is known that thermal leptogenesis can be associated with testable
($\mgr$-dependent) upper limits on the masses of the gluino and the
other gauginos.
Nevertheless, the $\tau_{\slepton}$ constraint from CBBN of $^6$Li is
not taken into account in any of these studies.
In fact, as mentioned in the Introduction, this $\tau_{\slepton}$
constraint disfavors thermal leptogenesis in constrained scenarios
with universal soft SUSY breaking parameters at
$\MGUT$~\cite{Pradler:2006hh,Pradler:2007is,Pradler:2007ar,Kersten:2007ab}.
Interestingly, in this work, it is this upper limit on
$\tau_{\slepton}$ that allows us to arrive in a model independent way
at a testable prediction of thermal leptogenesis that does not depend
on the gravitino mass $\mgr$.

As can be seen in Fig.~\ref{Fig:ProbingC}, already the conservative
limits imposed by $\tau_{\slepton}\leq 10^4\,\seconds$ show that
$\TRmax=10^9\,\GeV$ is associated with a gluino-slepton mass ratio of
$c<3$ in the collider-friendly region of $m_{\slepton}< 1~\TeV$.
Thus, this ratio favors a gluino mass that will be accessible at the
LHC. In this way, $c<3$ can be understood as a testable prediction of
thermal leptogenesis for phenomenology at the LHC.
Indeed, the realistic, less conservative constraints
$\tau_{\slepton}\leq 5\times 10^3\,\seconds$ and $\TR\geq 3\times
10^9\,\GeV$ point even to the exceptional case with a gluino that is
only slightly heavier than the slepton NLSP: $c\lesssim 1.5$.
If realized in nature, this will guarantee unique signatures at the
LHC.

\section{Conclusion}

The observation of a (quasi-) stable heavy charged slepton $\slepton$
as the lightest Standard Model superpartner at future colliders could
provide a first hint towards the gravitino LSP being the fundamental
constituent of dark matter. 
Under the assumption that the gravitino is the LSP and a long-lived
$\slepton$ the NLSP, we have shown that measurements of the slepton
mass $m_{\slepton}$ and its lifetime $\tau_{\slepton}$ can provide the
gravitino mass $\mgr$ and upper limits on the reheating temperature
$\TR$ after inflation, which can be tightened once the gluino mass
$m_{\gluino}$ is measured.

The conceivable insights into $\mgr$ are a direct consequence of the
SUGRA Lagrangian and of R-parity conservation.
Thus, they will be valid for any cosmological scenario provided the
$\gravitino$ is the LSP and a $\slepton$ the NLSP.
The possibility to probe $\TR$ at colliders results from the $\TR$
dependence of the thermally produced gravitino density and the fact
that this density cannot exceed the dark matter density.
Thereby, the given upper limits on $\TR$ rely not only on the SUGRA
Lagrangian and R-parity conservation but also on the assumptions of a
standard cosmological history and on (gauge) couplings that evolve at
high temperatures as expected from their standard MSSM renormalization
group running. 
Nevertheless, under these assumptions, the presented $\TR$ limits are
very robust since they have been derived in a model independent way
from a conservative lower limit on the (thermally produced) gravitino
density.

Prior to collider measurements of $m_{\slepton}$, $\tau_{\slepton}$,
and $m_{\gluino}$, primordial nucleosynthesis can be used to constrain
$\mgr$ and $\TR$. In particular, we have found that the constraint
$\tau_{\slepton}\lesssim 5\times 10^3\,\seconds$ from CBBN of $^6$Li
implies a truly conservative limit of $\TR \lesssim 3\times
10^9\,\GeV$ in the collider-friendly region of $m_{\slepton}\lesssim
1\,\TeV$ and in the limiting case of a small gluino--slepton mass
splitting at the weak scale: $c=m_{\gluino}/\mslepton\simeq 1$.
Indeed, with $c=7$---as obtained in scenarios with universal soft SUSY
breaking parameters at the scale of grand unification---our robust
conservative upper limit is $\TR \lesssim 10^8\,\GeV$ and thus as
severe as in scenarios with an unstable gravitino of mass $\mgr
\lesssim 5~\TeV$.

The $\tau_{\slepton}$ constraint from CBBN of $^6$Li has allowed us to
derive upper limits on $\TR$ that depend on $\mslepton$ and $c$ only.
In particular, we find that the condition for successful thermal
leptogenesis with hierarchical right-handed heavy Majorana neutrinos,
$\TR> 10^9\,\GeV$ ($3\times 10^9\,\GeV$), implies an upper limit of
$c<3$ ($1.5$) for $\mslepton<1~\TeV$.
This is a prediction of thermal leptogenesis that will be testable at
the upcoming LHC experiments.

\medskip

\begin{acknowledgments}
  I am grateful to Josef Pradler for valuable discussions and comments
  on the manuscript.
  This research was partially supported by the Cluster of Excellence
  `Origin and Structure of the Universe.'
\end{acknowledgments}
%

%
%
\end{document}